# Describing the complexity of systems: multi-variable "set complexity" and the information basis of systems biology

David J. Galas[1,2*], Nikita Sakhanenko[1], Alexander Skupin[2] and Tomasz Ignac[1,2]

[1]Pacific Northwest Diabetes Research Institute
720 Broadway
Seattle, Washington
98122, USA

[2]Luxembourg Centre for Systems Biomedicine

University of Luxembourg

7, Avenue des Hauts-Fourneaux

L-4362 Esch-sur-Alzette

Luxembourg

* corresponding author:  e-mail address:  dgalas@pnri.org

## Abstract

Context dependence is central to the description of complexity. Keying on the pairwise definition of "set complexity" we use an information theory approach to formulate general measures of systems complexity. We examine the properties of multi-variable dependency starting with the concept of interaction information. We then present a new measure for unbiased detection of multi-variable dependency, "differential interaction information." This quantity for two variables reduces to the pairwise "set complexity" previously proposed as a context-dependent measure of information in biological systems. We generalize it here to an arbitrary number of variables. Critical limiting properties of the "differential interaction information" are key to the generalization. This measure extends previous ideas about biological information and provides a more sophisticated basis for study of complexity. The properties of "differential interaction information" also suggest new approaches to data analysis. Given a data set of system measurements differential interaction information can provide a measure of collective dependence, which can be represented in hypergraphs describing complex system interaction patterns. We investigate this kind of analysis using simulated data sets. The conjoining of a generalized set complexity measure, multi-variable dependency analysis, and hypergraphs is our central result. While our focus is on complex biological systems, our results are applicable to any complex system.

**Introduction**

A living system, while invariably complex, is arguably distinguished from its non-living counterparts by the way it stores and transmits information. It is just this information that is at the heart of the biological functions and structures. It is also at the center of the conceptual basis of what we call systems biology, and is characterized by complexity that is both high in degree, involving a large number of components, variables and attributes, and difficult to characterize. The definition, and the conceptual context of what we call complexity is an ongoing discussion, but here we eschew the philosophical issues and focus on as simple, but precise, a description as we can. The conceptual structure of systems biology, we argue here, can be built around the fundamental ideas concerning the storage, transmission, and use of biological information, and descriptions of their collective complexity. Bio-information resides, of course, in digital sequences in molecules like DNA and RNA, but it is also in 3-dimensional structures, chemical modifications, chemical activities, both of small molecules and enzymes, and in other components and arrangements of components and properties of biological systems at many levels. The information depends simply on how each unit interacts with, and is related to, other components of the system. Biological information is therefore inherently context-dependent, which raises significant issues concerning its quantitative measure and representation. An important and immediate issue for the effective theoretical treatment of biological systems then is: how can context-dependent information be usefully represented and measured? This is important both to the understanding of the storage and flow of information that occurs in the functioning of biological systems and in evolution. It is also at the heart of the analysis of data extracted from complex systems of all kinds, including biological systems.

In a system with a number of variables, attributes and characters, one statement of the fundamental problem of definition and discovery of bio-information revolves around the question: how can we fully describe the joint probability density of the $n$ variables that define the system (as a function of time as well)? Characterization of the joint probability distribution is at the heart of describing the mathematical dependency among the variables. For many reasons, including important applications, this problem, has received a lot of attention over past decades, primarily focusing on binary relationships, and in drawing conclusions about multiple variable dependencies from these, as in copula theory, for example, and its central result, Sklar's theorem [14,15]. Here we provide a general formulation of the problem and a solution that deals directly with the dependency issue based on multi-variable information theory. We do not, in this work, address in any detail the significant problems of implementation of the computations implicated, their efficiency and properties. Rather we lay out the formulation of the approach, the theoretical structures, and with a few simulated examples illustrate their utility. We thereby provide a number of tools that are useful in the quest for the description of complex biological systems, and complexity in general.

**Results**

**a. Describing a Complex System**

It is often assumed that the networks that are at the heart of biological systems can be fully described by graphs, and most often by undirected graphs. The representations of relationships that graphs provide are rich indeed, but the complexity of biological systems (and many others) can go well beyond the ability of graphs to represent their full complexity. Even directions, weights, and other attributes assigned to edges can fall short of what is required. Graphs, made up of nodes, and edges connecting these nodes two at a time, are potentially complex mathematical objects, but are limited primarily in one way. A full description of a complex system often requires that a relationship among several components at once be described. For example, multiple external parameters like ionic strength and temperature, and many other variables, can together affect biological states. Biochemical reactions that involve more than two participating molecular partners are most common: $A + B \Leftrightarrow C$, involves three, and more complex reactions, like $A + B \Leftrightarrow C + D$ are common. Multiple, interacting transcription factors affecting the expression of genes, multiple proteins interacting closely together in protein complexes with a wide range of function, are other examples in biology.

Hypergraphs are sufficiently complex mathematical structures to describe the level of complexity required for a full description of the dependencies among the components, attributes or variables, in a biological system [4] as has been effectively argued. Undirected hypergraphs, which we will use here (generalization to directed hypergraphs is also possible), consist of nodes or vertices and edges, like a simple graph. Note that there remain significant attributes and parameters that cannot be represented as simple hypergraphs. The edges (we will term these hyperedges), however, may connect any number of vertices at once. Formally, a hypergraph, $HG=(V,E)$, consists of a set of vertices, $V=\{X_1, X_2, X_3, \ldots X_n\}$, and a set of subsets of $V$, we call edges, $E = \{\varepsilon_1, \varepsilon_2, \varepsilon_3, \ldots, \varepsilon_m \mid \varepsilon_i \subseteq V\}$. A simple graph, then, is a hypergraph with all of the hyperedges having a cardinality of 2. To illustrate, the connected hypergraph of nine vertices below has hyperedges of cardinality 2, 3 and 4.

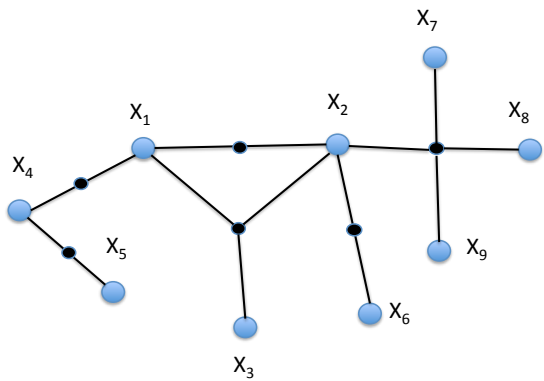

Figure 1a. A hypergraph of nine variables with hyperedges of order 2, 3 and 4.

The hyperedges consist of the subsets $\{X_2, X_7, X_8, X_9\}$, $\{X_1, X_2, X_3\}$, $\{X_2, X_6\}$, $\{X_1, X_2\}$, $\{X_1, X_4\}$, $\{X_4, X_5\}$. If we use hypergraphs to describe a system, we simply need to define the nature of the hyperedge, including the subset of nodes connected, weight, other edge attributes. If the vertices in a biological system were proteins, the hyperedge may describe, for example, the relations "forms a complex" or "regulates". A specific hyperedge could also connect a subset of another hyperedge of the hypergraph: as with $\{X_1,X_2,X_3\}$ and $\{X_1,X_2\}$, above. Note the similarity of a weighted hypergraph to a Markov random field.

There are, of course, many properties that need description to fully characterize a system, but one property essential with respect to quantitative variables could be described simply as "depends on", and its inverse "is independent of". We will focus here on dependence. We define the collective dependence of a subset of variables in the strict sense here that a variable, or attribute can be predicted only if all the other members of the subset are known. The notion of "independent of" is defined mathematically as the factorability of the joint density distribution. The notion that there is a dependence, or pattern present only in the entire subset of variable values, but not in any of its proper subsets, is fundamental, and is delineated well by hypergraphs. A node can, of course, be part of more than one hyperedge. Consider this simple example where the nodes have numerical values.

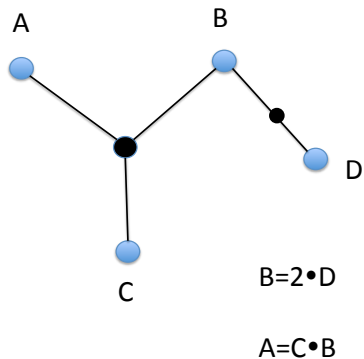

B=2•D

A=C•B

Figure 1b. A hypergraph with "overlapping" edges, and the functional dependencies among the variables corresponding to the nodes.

The two relationships in Figure 1b apply to B, a member of both edges, simultaneously. The number of ways that a set of variables can depend on one another is large, and the classification of these dependencies needs detailed consideration.  Looking at the two extremes, for example, what we might call "full dependency" among $n$ variables permits knowledge of all variables when only one is known, while the "collective dependency", defined above, is quite different and requires knowledge of all others to predict one.

In practice, in trying to decipher the complexity of a system we are often presented with a data set, values of a set of variables, and we wish to determine and then describe the dependencies among them.  In the reality of experimental measurements this is usually viewed statistically, and most often as a set of pairwise correlations.    *We are presented with the problem of estimating most accurately, given a data set with values of variables or attributes, the dependencies of the system, no matter how complex they may be.*   As we demonstrate in a later section, correlation functions, like Pearson and Spearman, since they are inherently pairwise, and have other limitations, like the assumption of relatively simple functional dependencies, are often inadequate.  If a data set was provided corresponding to the hypergraph depicted in Figure 1, how could we test, or infer the given structure?   More simply, given a data set of variables, is there a systematic way of inferring the hypergraph that represents that system?  This is the same as the challenge of discovering the structure of variable interdependence and representing it in a hypergraph.  To give this a quantitative context, let us add a "weight" to each edge that describes the "strength" or reliability of the dependency. We say that a weight of zero means there is no hyperedge, the variables are not dependent, and a weight near zero indicates a very weak one.    At the heart of this challenge is the realization that each variable acquires its meaning and relations from the context of the whole system, and the discovery of representation of this is hard.  In order to state the problem clearly we will revert here to a statistical definition of dependency (next section), which is well-matched to the data analysis aspects of our concerns.   Jakulin and Bratko [6] advocate the concept of interaction among attributes as a characterization of regularities or patterns among variables, that includes the multiple possibilities of conditional independence, which is similar to our approach.   In any case, a key representation of the dependency description is a hypergraph with marginal probability densities associated with vertices, and the hyperedges describing dependencies.

It is important to recognize the separation of the two fundamental problems of dependency, the problem of detection of dependency among variables, and the problems of determining the nature of the dependency. Simply stated this is the difference between detecting existence and estimating a function. These are very distinct problems, and while we address only the first of these in this paper, the significance of their separation is paramount. Note as well that both problems are completely distinct from the question of causality.

## b. Context-dependent measures: set complexity

Before grappling with full multivariable dependency we consider pairwise dependence only. We deal here with the problem of the information about one variable represented in another. The information perspective here is a useful one. The basic idea is illustrated by considering a set of bit strings, $\{x_i\}$. We can ask: what is the information in a given string in the context of the rest of the set of strings (for simplicity we assume the digital strings here are the same length.) Two useful concepts here are the mutual information between two strings, and the normalized information distance between these strings (this universal distance was defined by Li and Vitanyi [2] and shown to be a metric which does not require the lengths of the strings to be the same.) The distance between two strings $x$ and $y$, $d(x,y)$, is

$$d(x,y) \equiv \frac{\max(K(x \mid y), K(y \mid x))}{\max(K(x), K(y))}$$

(1a)

where $K(x|y)$ is the conditional complexity or conditional information between the strings. The values of $d$ lie in the interval $[0,1]$. We used the above quantities in [1] as defined either in the context of Kolmogorov complexity, or using the Shannon information formalism. While they are different ideas and are applied differently the concepts are conceptually interchangeable for our purposes, even though they are not quantitatively identical as has been carefully detailed previously [1]. The use of Kolmogorov complexity has advantages, like obviating the need to define a state space fully for each variable, but has the major disadvantage of being incomputable, but estimatable using compression algorithms. In this paper we use identical alphabets and the same numbers of data points for all variables considered together. We leave the discussion of the difference between these approaches to quantitating information for elsewhere.

We rewrite equation 1a, using information or entropies rather than Kolmogorov complexities, using the relation among the variable information, $H(X_1 | X_2) = H(X_1, X_2) - H(X_2)$.

$$d(X_1, X_2) = \frac{\max[H(X_1, X_2) - H(X_1), H(X_1, X_2) - H(X_2)]}{\max[H(X_1), H(X_2)]} \qquad (1b)$$

The value of $d$ when the two variables are entirely dependent (they are functions of each other) is zero since the joint entropy is equal to any of the two marginal entropies. At the opposite extreme, when the two variables are entirely independent (are independent random variables, for example), the value of $d$ is also easy to determine. In this case the single variable entropies are the same (since the length of the strings or the range of the variables is the same), and the joint entropy is the sum of the two, thus for <u>complete independence</u> of these two variables we have

$$d(X_1, X_2) = \frac{H(X_1) + H(X_2) - H(X_2)}{H(X_1)} = 1 \qquad (1c)$$

As we have proposed [1] the key to defining a real context-dependent measure for a member of the set is to impose constraints that have the effect of minimizing the contributions of both redundancy, and randomness. We seek a heuristic compromise that for two variables at a time represents the following two constraints (Notation: we use lower case to indicate a string or data, and upper case to indicate a variable): 1) if the set already contains a string $x$ identical to the one being considered, $y$, so that $d(x,y)=0$, then y adds no new knowledge to the set [1];

2) if $y$ is random (defined strictly only when the length of the string increases without bound), then $y$ also adds nothing to the complexity of the set [1]. These criteria are fulfilled approximately if we define the information contribution of an element $x_j$ to the set $S=\{x_i|i = 1,2,\ldots,N\}$, $k(x_j|S)$, in terms of the pair-wise information distance $d(x,y)$ in (1a):

$$k(x\,|\,S) \equiv \frac{1}{N-1}\sum_{y\in S} K(y)d(y,x)(1-d(y,x)) \qquad (2)$$

in which $K(y)$ is the information measure of an individual element $y \in S$ [1]. We then defined the complexity of the entire set of strings, the "set complexity", $\Psi$, [1] as

$$\Psi(S) \equiv \frac{2}{N(N-1)}\sum_{all-pairs} \max[K(x_i),K(x_j)]d(x_i,x_j)(1-d(x_i,x_j)) = \sum_j k(x_j\,|\,S) \qquad (3a)$$

and if we order the variables $\{x_i\}$ by the increasing magnitude of $K(x_i)$ then

$$\Psi(S) = \frac{2}{N(N-1)}\sum_{all\_pairs} \phi_{i,j} = \left\langle \phi_{i,j} \right\rangle \;.\; i>j \qquad (3b)$$

where $\phi_{i,j} = K(x_i)d(x_i,x_j)(1-d(x_i,x_j))$ . If we use the entropy of the probability distribution density as the information measure, as we will in the rest of this paper, we indicate this as $K(x) = H(x)$.

### c. Describing multi-variable dependence

The description of a complex system would be severely limited if we restricted ourselves to considering only pairs of variables or functions. It is therefore easily argued that we must define measures for multiple variables considered together rather than by pairs.

The concept of "interaction information" [8,9] was proposed long ago, and was used previously by us to optimize a binning process so as to minimize bias and lose a minimum of information in the process [11]. Interaction information is essentially a multi-variable generalization of mutual information [5]. For two variables the interaction information is equal to the mutual information and to the Kullback-Leibler divergence of the joint to single probability densities of these two variables. Interaction information (essentially the same as co-information as defined by [10]) expresses a measure of the information shared by all random variables from a given set [5-7]. For more than two variables it has properties distinct from mutual information, however, including potentially negative values, but it remains symmetric under permutation of variables and has been used in several applications to date.

We first extend the interaction information from two- to three-variables. The three-variable interaction information, $I(X_1, X_2, Y)$ can be thought of as being based on 2 predictor variables, $X_1$ and $X_2$, and a target variable, $Y$ (there is actually nothing special about the choice of a target variable since $I$ is symmetric under permutation of variables, but this will be important in later considerations.) The three-variable interaction information can be written as the difference between the two-variable interaction information, with knowledge of the third variable, and the two-variable quantity without that knowledge:

$$I(X_1, X_2, Y) = I(X_1, X_2) - I(X_1, X_2 | Y) \qquad (4)$$

where $I(X_1, X_2)$ is mutual information, and $I(X_1, X_2 | Y)$ is conditional mutual information, given $Y$. Note that if the additional variable is independent of the others the interaction information is zero. When expressed entirely in terms of marginal entropies we have the expression:

$$I(X_1, X_2, Y) = H(X_1) + H(X_2) + H(Y) - H(X_1, X_2) - H(X_1, Y) - H(X_2, Y) + H(X_1, X_2, Y) \quad (5)$$

$H(X_i)$ is entropy of a random variable $X_i$, and $H(X_{k1}, \ldots, X_{km})$, $m \geq 2$, is a joint entropy on a set of $m$ random variables. The symmetry under variable permutation we mentioned above is apparent from equation 5.

We can write the interaction information in terms of sums of marginal entropies according to the inclusion-exclusion formula [6,7], which is the sum of the joint entropies of the sublattice of $v = \{X_1, X_2...X_n\}$ as described by Bell [7]. We have,

$$I(v) = \sum_{\tau \subseteq v} (-1)^{|\tau|+1} H(\tau)$$

(6a)

where the exponent, $|\tau|$, is the cardinality of the the subset $\tau$. Note that there is also a symmetrical formula (a form of Möbius inversion) defining the joint entropy in terms of the interaction information of the subsets.

$$H(v) = \sum_{\tau \subseteq v} (-1)^{|\tau|+1} I(\tau)$$

(6b)

Simplifying the notation further, we note the first two interaction information expressions (the subscripts refer to the index of variables):

$$I_{12} = H_1 + H_2 - H_{12}$$
$$I_{123} = H_1 + H_2 + H_3 - H_{12} - H_{13} - H_{23} + H_{123}$$

(7)

Note that the number of terms grows as a power of the number of variables.

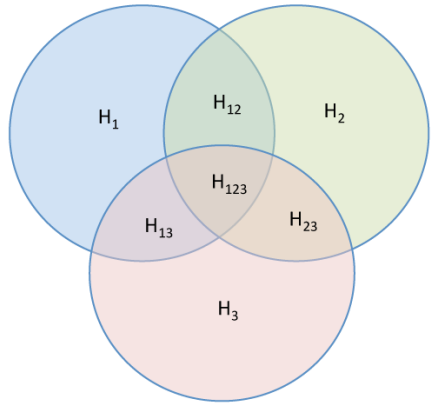

Figure 2. An illustration of the relationships between the terms in Equation 5, where the external outline encompasses the area represented by $I_{123}$.

Using equation 1b we can write

$$1 - d(X_1; X_2) = \frac{H(X_1) + H(X_2) - H(X_1, X_2)}{\max[H(X_1), H(X_2)]} = \frac{I(X_1, X_2)}{\max[H(X_1), H(X_2)]} \tag{8}$$

Thus, we have proven the following theorem.

**Theorem 1**: The pairwise <u>set complexity</u>, as defined in equation 3, is simply related to the pairwise interaction information in a set of variables as a normalized expectation value of the larger of two terms over all pairs of variables:

$$\Psi(S) = \left\langle \frac{\max[H(X_i \mid X_j), H(X_j \mid X_i)] \cdot I(X_i, X_j)}{\max[H(X_i), H(X_j)]} \right\rangle \tag{9}$$

The key property of this measure, $\Psi$, is that at the two extremes – when either all variables are identical, (or fully pair-wise dependent), or all variables are independent (or random with respect to one another) – the set complexity is zero. Since "set complexity" can be viewed as an average over all pairs of strings, variables or functions, every well-defined subset of a larger set will also have a well defined $\Psi$. The original motivation for the requirement for zeros at the extremes was the resolution of two problems with the idea of biological information and the product was chosen to force the measure to have zeros at these extremes [1]. While we know that there are many other ways to get zeros at the extremes, the direct relationship of this simple form with the interaction information suggests that our original heuristic, form was actually a natural and fortunate choice.

We begin the construction of our measure with the "interaction information" for multiple variables as defined in equations 4 and 5. Using this concept for a set of $n$ variables, $v_n = \{X_1, X_2, X_3 \dots X_n\}$, we define the "differential interaction information" for this set of variables, $\Delta$, as the change in the interaction information between sets that differ only by the addition of one variable. Thus, if $v_n$ is obtained from $v_{n-1}$ by the addition of $X_n$, we have

$$\Delta_{X_n}(v_n) \equiv \Delta_n \equiv [I(v_n) - I(v_{n-1})] = -I(v_{n-1} \mid X_n) \tag{10}$$

The last equality comes from the recursion relation for the interaction information, equation 4. The added variable, $X_n$, we will call the "target variable", with respect to the variable set $\nu_{n-1}$. The differential interaction information is equivalent to the conditional interaction information, which for the three variable case is equivalent to the conditional mutual information. It is easy to see that the differential interaction information is zero if $X_n$, the target variable, is independent of any of the variables in the set $\nu_{n-1}$, that are independent of one another. We show that if there is collective dependency of the variable set the differential measure in equation 10 will be non-zero.

We can write the differential interaction information in terms of the marginal entropies. If $\nu_n = \{X_1, X_2, X_3, \ldots X_n\}$, and $\{\tau_n\}$ are all the subsets of $\nu_n$ that contain $X_n$ (not all subsets) then

$$\Delta_n = \sum_{\{\tau_n\}} (-1)^{|\tau_n|+1} H(\tau_n) \qquad (11)$$

The notation is simplified here so that $\Delta_n$ means an $n$-variable measure where the $n$th variable is the "target variable." Where this may be ambiguous we will indicate more parameters.

For 3 and 4 variables we can write out equation 11 as (indicating the number variables by the subscript of delta)

$$\Delta_3 = I_{123} - I_{12} = H_3 - H_{13} - H_{23} + H_{123}$$
$$\Delta_4 = I_{1234} - I_{123} = H_4 - H_{14} - H_{24} - H_{34} + H_{124} + H_{134} + H_{234} - H_{1234}$$

$$(12a)$$

where the subscripts of the marginal entropies on the right indicate the variable indicies. Note that we can show with a little algebra that the three variable measure in equation 12a is simply the conditional mutual information. The mutual information, and its conditional for $X_1$, $X_2$ and $X_3$ are

$$I(X_1, Y_2) = H(X_1) + H(X_2) - H(X_1, X_2)$$
$$I(X_1, X_2 \mid X_3) = H(X_1 \mid X_3) + H(X_2 \mid X_3) - H(X_1, X_2 \mid X_3)$$

By writing out the conditional entropies in terms of marginal entropies, and adopting our previous simplified notation we see that the conditional mutual information and the differential interaction information for three variables are the same.

$$I(X_1, X_2 \mid X_3) = H_{13} + H_{23} - 2H_3 - H_{123} + H_3$$
$$= H_{13} + H_{23} - H_3 - H_{123} = -\Delta_3$$

(12b)

Since there is no definition of interaction information for one variable, the definition of the differential for two variables, $\Delta_2$, cannot be directly defined by extention from equation 11. This extension has an unexpected significance and we will return to consider it shortly. Note that the number of terms grows as a power of the number of variables minus one.

We now use the differential interaction information, $\Delta_n$, to define a generalized form of the set complexity, $\Psi$, for an arbitrary number of variables. The information distance measure (2 variables) has the property that as the two strings converge to identity the normalized mutual information increases to 1, and as they become entirely independent (random) it goes to zero. The distance goes from zero (identity) to one (independence.) In the case of three variables, where we will use $\Delta_3$, it is significantly more subtle. When $\Delta_3$ is positive it indicates "redundancy" – moving towards identity, while when it is negative it indicates "synergy" or some functional dependency of the three variables that is not identity. This difference in sign is significant. Specific examples help illustrate this difference.

*Example 1.* Consider three random variables, $X_1$, $X_2$, and $X_3$. Let us evaluate $\Delta_3$ for the case when the variables are all independent. Denote this $\Delta_3^0$:

$$\Delta_3^0 = -H_3 - H_1 - H_2 + H_1 + H_2 + H_3 = 0 \qquad\qquad (12c)$$

It is easy to show that $\Delta_i$ is zero for all numbers of variables, $i$, if all variables are independent since the joint marginal entropies become additive single entropies.

*Example 2.* Let us now evaluate $\Delta_3$ when $X_3$ depends on $X_1$ and $X_2$ together, but on neither singly. The expression then becomes

$$\Delta_3 = H_3 - H_{13} - H_{23} + H_{123} = H_3 - H_1 - H_3 - H_2 - H_3 + H_{123} = -H_1 - H_3 - H_2 + H_{123}$$

and because of the assumed dependency,

$$H_{123} < H_1 + H_3 + H_2,$$

so $\Delta_3 < 0$. Thus, since $X_3$ is the "target variable" (the asymmetric variable in the definition of $\Delta_3$), in this case $\Delta_3$ works as an indicator of three-variable dependence. Notice that the symmetry of the interaction information does not carry over to the differential interaction information. Since we are searching for a measure that reliably distinguishes between two-variable and three-variable dependence of all kinds we need our measure to vanish in the presence of only two variable dependencies. Note that the arguments to follow are exactly true only in the limit of the number of values of the variables (data) increasing without bound. The differential interaction information above actually becomes zero only in some cases, as we can see from two more examples.

*Example 3.* Suppose $X_1$ and $X_3$ are independent of each other, as are $X_1$ and $X_2$, but $X_2$ and $X_3$ are dependent. In this case $H_{123}=H_1+H_{23}$ (in the limit), so we have

$$\Delta_3 = H_3 - H_{13} - H_{23} + H_{123} = H_3 - H_1 - H_3 - H_{23} + H_1 + H_{23} = 0$$

Examples 2 and 3 show that if the pair-wise dependency includes $X_3$ then three-variable dependency is needed to get a non-zero $\Delta_3$. This is the behavior we want from our measure, that parallels the two variable measure properties, so it works in these cases.

*Example 4.* Now suppose $X_1$ and $X_3$ are independent of each other as are $X_2$ and $X_3$, but $X_1$ and $X_2$ are dependent. In other words, the target variable $X_3$ is no longer dependent on the other two variables. The same arguments apply as in example 3, however $\Delta_3$ is seen to be non-zero. Example 4 then shows that this measure, $\Delta_3$, fails to be three-variable-specific.

It is clear then that we need something else. The measure we want is a non-zero quantity for a subset, $\tau$, of $m$ variables, <u>only</u> if there is mutual dependency among the elements of the subset, $\tau$.

The differential interaction information, $\Delta$, we have described thus far in equation 11 is based on the specification of a variable we called the "target variable" within the set of variables. The differential is defined as the change that results from addition of this target variable, and is therefore asymmetric under permutation of the variables. Since we are asking to detect *fully cooperative dependence* among the variable set, we require a useful measure to be symmetric for that set. A more general measure emerges by a simple construct that restores symmetry. If we multiply $\Delta$'s for a given variable subset with all possible choices of a target variable, the resulting measure will be symmetric for all the variables in the set. It provides a general definition that is functional and straightforward. To be specific, we define the symmetric measure (with normalization) as

$$\overline{\Delta}_n = \overline{\Delta}(v_n) \equiv (-1)^n \prod_{i=1}^{n} [I(v_n) - I(v_n \setminus \{X_i\})] \qquad (13)$$

where the product is over the choice, $i$, of a target variable relative to $v_n$, $n>2$, a simple permutation. The difference terms in the bracket in equation 13 are between the interaction information for the full set $v_n$ (first term) minus the interaction information for the same set missing a single element (the target variable - second term.) We define this measure only for $n>2$ because the differential interaction information for $n=2$ is as yet undefined.

For three variables this expression is (simplifying the notation again)

$$\overline{\Delta}_3(1,2,3) = (-1)^3 (H_1 - H_{12} - H_{13} + H_{123})(H_2 - H_{12} - H_{23} + H_{123})(H_3 - H_{13} - H_{23} + H_{123})$$

$$(14)$$

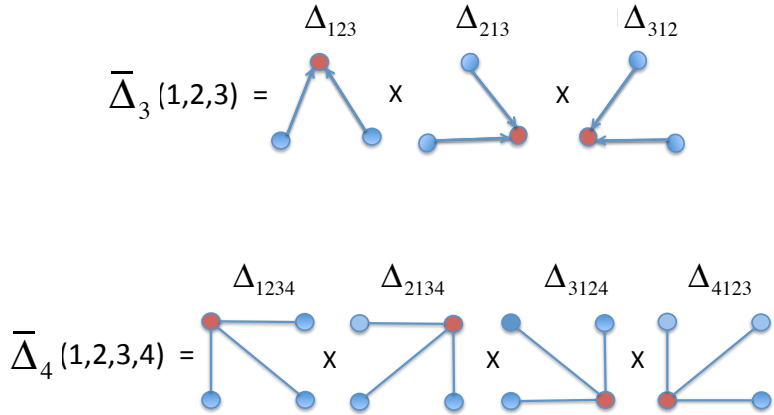

Figure 3. A diagram illustrating the factors in the symmetrized product for the three- and four-variable cases. The $\Delta$'s in this diagram signify the factors in the symmetrized product. The upper one, for example signifies the terms in eqn 13. The first subscript index of each $\Delta$ is the target variable.

The advantage of this measure is that $\overline{\Delta}$ is non-zero only if there is a "collective" dependency, with all three variables involved. In other words, this measure has the extremely useful property that it always vanishes unless <u>all</u> variables in the subset are interdependent. This can be used to allow us to discover and represent exact variable dependencies, and to define complexity in interesting ways.

To illustrate these differences among the interaction information, $I$, the asymmetric differential interaction information, $\Delta_i$, and the symmetric product form, $\overline{\Delta}$, we have formalized the definitions and calculated the values for some extreme dependencies for three variables, shown in table 1. In a later section we will discuss dependency in terms of "well-behaved functions", and show some examples.

Consider variables $X_1$, $X_2$, and $X_3$.

**Definition 1.** Full Independence (*FI*)

$$X_i \perp X_j, \forall i \neq j$$

**Definition 2.** Full Dependence (*FD*)

$$X_i \Rightarrow X_j, \forall i, j$$

**Definition 3.** Two variable Dependence (*2vD(i,j)*)

$$X_i \Rightarrow X_j$$
$$X_i \perp X_k$$
$$X_j \perp X_k$$
$$i \neq j \neq k$$

**Definition 4.** Collective dependence (*CD(i)*)

$$(X_j \,\&\, X_k) \Rightarrow X_i$$
$$X_i \perp X_j, \;\; X_i \perp X_k, \, X_j \perp X_k \quad \forall j,k \neq i$$

Note that the statements of independence in definition 4 include the cases where $X_i$ is independent of the other single variables, but dependent on two of them together.

| | | $I_3 = H_1 + H_2 + H_3 - H_{12} - H_{13} - H_{23} + H_{123}$ | $\Delta_3 = H_3 - H_{13} - H_{23} + H_{123}$ | $\bar{\Delta}_3 = (H_3 - H_{13} - H_{23} + H_{123}) \times (H_2 - H_{12} - H_{23} + H_{123}) \times (H_1 - H_{12} - H_{13} + H_{123})$ |
|---|---|---|---|---|
| *FI* | | 0 | 0 | 0 |
| *FD* | | $H_{123} = H_1$ | 0 | 0 |
| *2vD* | $(X_1, X_2)$ | 0 | $-H_1 = -H_2$ | 0 |
| | $(X_1, X_3)$ | 0 | 0 | 0 |
| | $(X_2, X_3)$ | 0 | 0 | 0 |
| *CD* | $(X_2 \& X_3) \Rightarrow X_1$ | $-H_1 = -H_2 - H_3$ | $-H_1 = -H_2 - H_3$ | $(-H_1)^3$ |
| | $(X_1 \& X_3) \Rightarrow X_2$ | $-H_2 = -H_1 - H_3$ | $-H_2 = -H_1 - H_3$ | $(-H_2)^3$ |
| | $(X_1 \& X_2) \Rightarrow X_3$ | $-H_3 = -H_1 - H_2$ | $-H_3 = -H_1 - H_2$ | $(-H_3)^3$ |

Table 1: Expressions (in simplified notation) and specific values of these quantities for various cases of dependence (see the formal definitions earlier).

The above table illustrates why $\overline{\Delta}$ is needed: this measure goes to zero for the first thee cases above, but is non-zero when there is a collective, three-variable dependency (fourth case.)

We can complete the theory now by providing a consistent definition for $\overline{\Delta}_2$ which connects this measure to the two variable set complexity (equations 3 and 9.) Since we cannot express it as a differential interaction information we define $\overline{\Delta}_2$, referring back to the key limiting values of set complexity, simply as

$$\overline{\Delta}_2 = -(H_1 + H_2 - H_{12})(H_1 - H_{12}) \tag{15}$$

Note that this expression has the property of going to zero for either complete independence (first factor goes to zero) or full dependence (second factor goes to zero) of the two variables, consistent with the properties of the expressions for more variables. It is symmetric under exchange of the two variables. Compare this with $\overline{\Delta}_3$ shown in equation 14. Now since $\phi_{ij} = H(X_i)d(X_i,X_j)(1 - d(X_i,X_j))$ from equation 8a, and referring to the definition of $d(X_1,X_2)$ in equation 1, we see easily by substitution that

**Theorem 2**: For two variables $X_1$ and $X_2$, if $H_1 < H_2$, and with the definition in equation 15, we have

$$\phi_{12} = \frac{1}{H_2}(H_1 + H_2 - H_{12})(H_{12} - H_1) = \frac{\overline{\Delta}_2(12)}{H_2} \tag{16}$$

Thus, the "set complexity" of a set of $N$ variables $v = \{X_i\}$ in terms of $\overline{\Delta}$ is

$$\Psi(v) = \left\langle \phi_{ij} \right\rangle = \frac{2}{N(N-1)} \sum_{all-pairs} \frac{\overline{\Delta}_2(ij)}{H_{ij}} = \left\langle \frac{\overline{\Delta}_2(ij)}{H_{ij}} \right\rangle \tag{17}$$

### d. Multi-variable Complexity Measures: generalizing set complexity

The remaining question is how we can use the multi-variable dependencies derived from any set of variables or attributes to describe the complexity of the full system they define. By analogy with the previous, two variable, "set complexity" measure, $\Psi$, we can use equations 8-10 to define a new, multi-variable class of set complexity measures. A complexity measure is an expectation value over the full system of a quantity defined on subsets of variables (in the case of the original set complexity these subsets are pairs.) When we examine the range of values of $\overline{\Delta}$ from complete independence of all variables to complete dependence, we discover an interesting property of the class of $\overline{\Delta}$ measures. From table 1 (column 5), illustrating the values of $\overline{\Delta}$ for different types of dependence, we see that $\overline{\Delta} = 0$. Full dependence here means knowledge of one variable gives us knowledge of all variables. For full dependence among three variables we have $H_{12} = H_1 = H_2$ and $H_{123} = H_1 = H_2 = H_3$. By calculating the product it can easily be seen that all terms cancel and $\overline{\Delta} = 0$. The interesting property of this measure for three variables is that it has zeros at both extremes (complete independence and full dependence.) That this property holds for an arbitrary number of variables provides a powerful result that we now present as a theorem.

**Theorem 3.** Given a set of variables $v = \{X_1, X_2, X_3, \dots X_{n+1}\}$ consider three cases: *a*) the variables are <u>independent</u>, and *b*) the variables are <u>fully dependent</u>, and c) the variable Xn+1 is independent of all the other variables. The value of $\overline{\Delta}$ in all three cases, *a*) , *b*) and c) is zero.

**Proof:**

Choose one variable from the set of *n+1* variables and call it $X_k$. Let $v_{n+1} = \{X_1, X_2, X_3, \dots X_{n+1}\}$, and $\{\tau_k\}$ denote all the subsets of $v_{n+1}$ that contain $X_k$. Consider the corresponding $\overline{\Delta}(v_{n+1})$, which has a factor $\Delta_k$ where $X_k$ is a target variable. From equation 11

$$\Delta_k = \sum_{\{\tau_k\}} (-1)^{|\tau_k|+1} H(\tau_k)$$

<u>Case *a*</u>: For complete independence the marginal entropies of m variables ( $|\tau_k| = m$ ),

$$H(\tau_k) = \sum_{X_i \in \tau_k} H(X_i) \tag{18}$$

By summing the coefficients of $H(X_i)$ for all $X_i \in \tau_k$ we see that the alternating signs of the contributing tuples for the marginal entropies leads simply to

$$\sum_{p=0}^{m-1} (-1)^p \binom{m-1}{p} = (1-1)^{m-1} = 0 \text{ for non-target variables,}$$

$$\sum_{p=0}^{m} (-1)^p \binom{m}{p} = (1-1)^{m} = 0 \text{ for the target variable } X_k,$$

so the theorem holds for case *a*.

<u>Case b:</u> For complete dependence it is even simpler, because we have

$$H(\tau_k) = H(X_k)$$

and we are able to reduce the entire expression to a sum of coefficients of $H(X_k)$. The sum adds to zero as in case *a*.

<u>Case c:</u> In this case we split the set $\{\tau_k\}$ into two non-overlapping subsets, $\{\hat{\tau}_k\}$ containing the independent variable $X_{n+1}$ and $\{\tilde{\tau}_k\}$ without $X_{n+1}$. The marginal entropy of each subset $\hat{\tau}_k$ is simply $H(\hat{\tau}_k) = H(X_{n+1}) + H(\tau'_k)$, where $\tau'_k$ is the subset of all variables in $\hat{\tau}_k$ except $X_{n+1}$. Then, applying this to the expression of $\Delta_k$, it is easy to see that entropies $H(\tau'_k)$ cancel out with entropies $H(\tilde{\tau}_k)$ and the coefficients of $H(X_{n+1})$ sum to zero, and therefore $\overline{\Delta}$ is zero. This proves the theorem.

Thus, as the equivalent of the pair-based "set complexity" we previously defined and used [1], we now define a general, multi-variable set complexity, $\Phi$, for the set of variables, *v*

<u>Definition:</u>

$$\Phi(v) \equiv \left\langle \frac{\overline{\Delta}(\tau)}{H(\tau-1)^{|\tau-1|}} \right\rangle \tag{19a}$$

where the expected value is over all possible subsets, $\tau$, of the variable set, and $H(\tau-1)$ is the maximum marginal entropy across all proper subsets of $\tau$ obtained by removing one variable. Writing it in detail, where $M$ is the number of subsets in $v$,

$$\Phi(v) = \frac{1}{M} \sum_{\tau \subseteq v} \frac{\overline{\Delta}(\tau)}{H(\tau-1)^{|\tau-1|}} \qquad (19b)$$

The above formula can be used to characterize the complexity of any subset of variables as well, of course, and therefore the components of the sum represent the hypergraph form of the set complexity represented by $\Psi$ for ordinary graphs. For two variables this expression reduces to the previous definition of $\Psi$ in reference [1]. Since the complexity is generally not well represented by a single number we can better characterize the complexity as the set of components of equation 19b, $\{\varphi(\tau) \mid \tau \subseteq v\}$, for all subsets of $v$ where

$$\varphi(\tau) \equiv \frac{\overline{\Delta}(\tau)}{H(\tau-1)^{|\tau-1|}} \qquad (19c)$$

This completes the generalization of the set complexity concept for an arbitrarily large set of variables, and provides a means of defining the "weights" of dependency in the hypergraph describing this system. Note that Theorem 3 provides a much more complete solution to the problems of biological information presented in reference [1] since it accounts for full dependence in a much more complete fashion.

### e. Describing dependency

We have discussed and formalized several limiting cases of dependency: collective dependency, full dependency, and full independence (see Definitions 1-4). We can add to the nuance of dependency by defining a more intuitive and rigorous dependency spectrum using functions that relate the variables in $v_n$, which will help us understand the relationships between the variables, entropies and the structure of the induced hypergraphs.

We begin with a definition.  A function, $f$, of a set of variables, $\tau$, is called "well behaved" if and only if the equation $f(\tau) = 0$ can be solved for <u>any</u> $X \in \tau$, to yield a function g, such that $x = g(\tau')$, where $\tau'$ is the set missing $x$.   For this to hold and if we are to have a single solution, $X$, then clearly $f(\tau)$ must be monotonic in each variable.  The dependency among the variables of the set $\tau$ can then be described by the set of all non-zero functions on all subsets of $\tau$: $\{f : f(\sigma) = 0,\ \sigma \subseteq \tau\}$.   For a well-behaved, monotonic function we need to specify all but one variable value to determine the value of the last one.  Keep in mind, however, that for many functions we expect to encounter the solutions, $X$, will not be unique.  The number of solutions is a key factor.

For the moment consider only functions with unique solutions so that the dependencies among the variables in the set $\tau = \{X_i\}$ are fully described by the set of all of the well behaved, monotonic functions

$$F = \{f_i(\eta_i) = 0, \forall \eta_i \subseteq \tau\}.$$

Let's look at some examples of functions that define dependencies.  Consider only pair-wise functions; that is, assume that there are only non-zero functions for the set that relate two variables at a time.  Then the functions $\{x_i = g_{ij}(x_j)\}$ define the edges of a graph of the set in which all non-zero functions correspond to an edge.   In this case, if the edges in a graph connect several nodes into a connected path, specifying one value determines the values of all the variables in the path since the functions allow successive solutions across the entire path.  We see that the dependence of this set of variables is <u>"full dependence"</u> if and only if the graph is connected.  This means that knowledge of one variable value defines the values of all the rest.  This can be seen to hold because all the variables are connected by a chain of solvable equations with unique solutions.  Since this property allows us to derive a functional relationship between any two variables on the path we infer the following result.

**Theorem 5:** If a dependence graph for a set of well-behaved, monotonic functions with pair-wise defined edges is connected it is always a complete graph.

We can define this completeness as <u>full dependence</u> (as in Definition 2 of table 1). The connected components of any dependence graph of this kind are always complete. This is not particularly complex in most cases although it does allow us to divide a set of variables into dependent subsets. A more interesting and realistic situation arises if we are able to infer a dependency from data and also attach a statistical measure to each dependence, a probability associated with each solution. In the general case when there are multiple solutions to the inverted functions we might ascribe equal probabilities to each, or, in the presence of noise or uncertainly, probabilities that reflect this lack of certainty. Suppose that there is a linear connectivity pattern (a path) with identified functions and probabilities. If the connected variables are serially indexed, and the probability of each function connecting two variables is $p_{j-1,j}$. Then, if we have a good value for variable $X_1$, the functions can be used to find values of all the other variables, but the probability associated with the final variable $X_n$ will be given by

$$prob(x_n) = \prod_{j=2}^{n} p_{j-1,j}$$

If, for example, there were two solutions for each function a reasonable, unbiased assumption would be to assign a probability of one half to each. Thus, the above product would be $\frac{1}{2}^n$. In the case of pairwise functions we can see that the propagation of values along a path, as described above, will be associated with decreasing probabilities of accuracy and therefore the completeness described in Theorem 5 becomes a more subtle property for these functions.

For non-pairwise functional dependencies the situation is more complex and interesting. For any subset of variables, if there is a single well-behaved function describing the relationship among the variables then the hyperedge is defined by this function. While the determined or known segments of a hypergraph are defined by these dependencies, we usually do not know these functions exactly if we know them at all.  The challenge is to determine the weight of an edge (including the level of confidence in the existence of such a function).   To our knowledge there is no simple equivalent of Theorem 5 for hypergraphs.  If a function is not monotonic in each variable the dispersion of probabilities will grow with the number of solutions.

This raises the issue of how the weights of the edges in a dependence hypergraph are related.   It is easy to find examples of dependency that are consistent with only a certain configuration of edges.  Another form of this question or issue might be addressed by a topic called the "interaction of dependency hypergraph edges."   There are a number of relationships that can be derived that represent the complexity that results from relaxation of the premise of Theorem 5.  This topic is beyond the scope of this paper and will be addressed elsewhere.

## Examples of Data Analysis

To illustrate the practical uses of the theory we analyze sets of simulated data generated by prescribed dependencies.  We have generated multiple data sets of five variables with different degrees of variable dependence:  no dependence (all are independent random variables), two dependent variables (one variable is dependent on one of the other independently generated random variables), and three variable dependence.  The details of the generation of these data sets are provided in the Appendix.  In addition, we have generated other sets of six variables including four-variable dependence.   In each of these cases, 1000 multivariable points were generated, and we have calculated the usual pair-wise correlations between variables (Pearson correlation and Spearman rank correlation).  Details are provided in the Appendix.  Example results are presented in the next set of figures.

The values of the measures indicated in Figure 4 illustrate the use of interaction information, and the power of the symmetrized differential interaction information $\overline{\Delta}$ to detect the three-way dependence of variables over the pairwise effects. In these examples, while the differential interaction information is suggestive of some kind of dependence, it is only the $\overline{\Delta}$ that definitively indicates the three-way dependence, which it does strongly as shown in figure 4b.

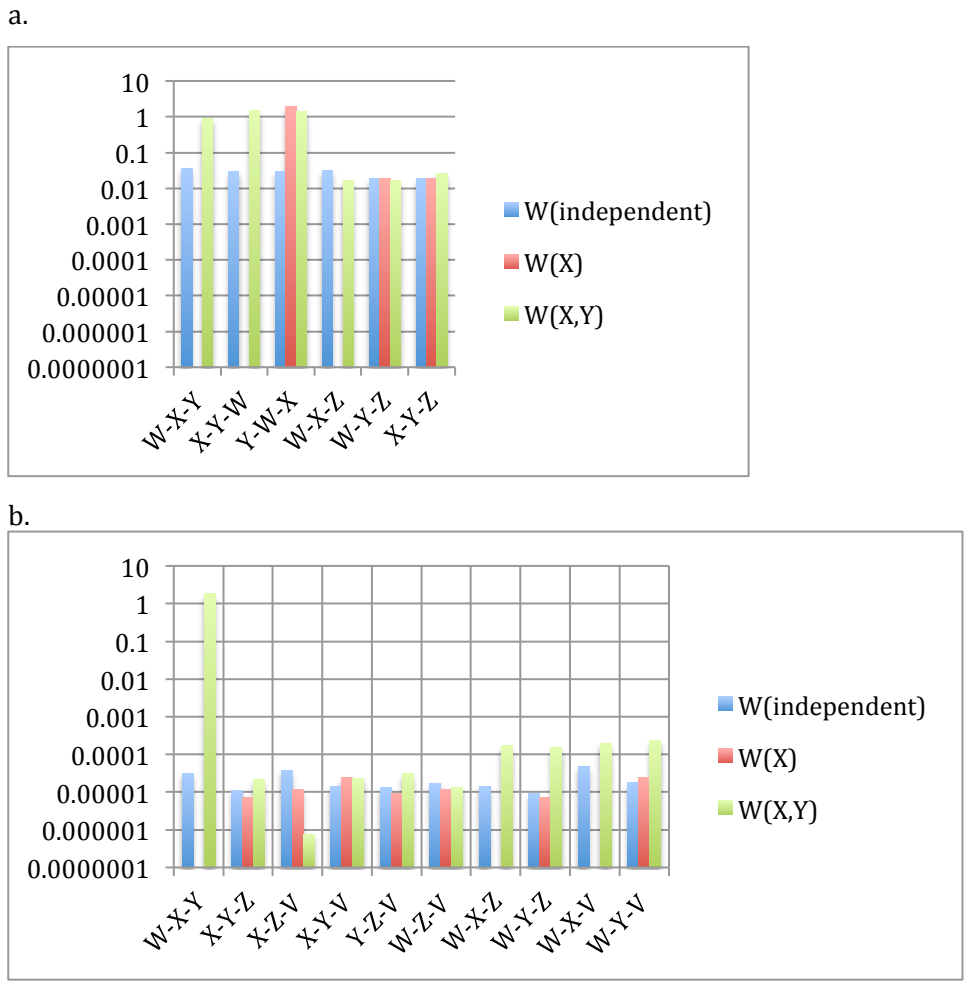

Figure 4. Information theory-based measures for three variables are shown for the simulated data (1000 points, 5 variables – see Appendix), as indicated. a. Differential interaction information in which the first variable indicated in the X-axis label is the target variable, and b. the symmetrized differential interaction information, $\overline{\Delta}$. The functions W(X) and W(X,Y) are complicated functions designed specifically to yield statistical correlations that are comparable to those of independent random variables (they are defined and illustrated in the Appendix and the correlations shown.) W(independent) represents that case where all variables are independent.

To illustrate the power of $\overline{\Delta}$ we further tested it by looking for four different variable dependencies in four data sets of six variables. We calculated the four variable $\overline{\Delta}$ for all these data sets and the results, illustrated in figure 5, show that the four variable $\overline{\Delta}$ clearly picks out the four variable dependency. In this case W is determined entirely by the values of X, Y, and Z, but not by V or U, where all five of these latter variables are independent random variables. Note that, since all variables take on integer values 0 to 3, there are 64 distinct sets of values for the triplet of variables, X, Y, and Z, while the function W(X,Y,Z) takes on only four integer values 0 to 3. The function is therefore far from monotonic, many values of {X,Y,Z} map into the same value of W. Nonetheless, the dependency is clearly indicated by the symmetric measure (the purple bar in figure 5. This mapping of variables onto W is illustrated in the Appendix (figure A2.)

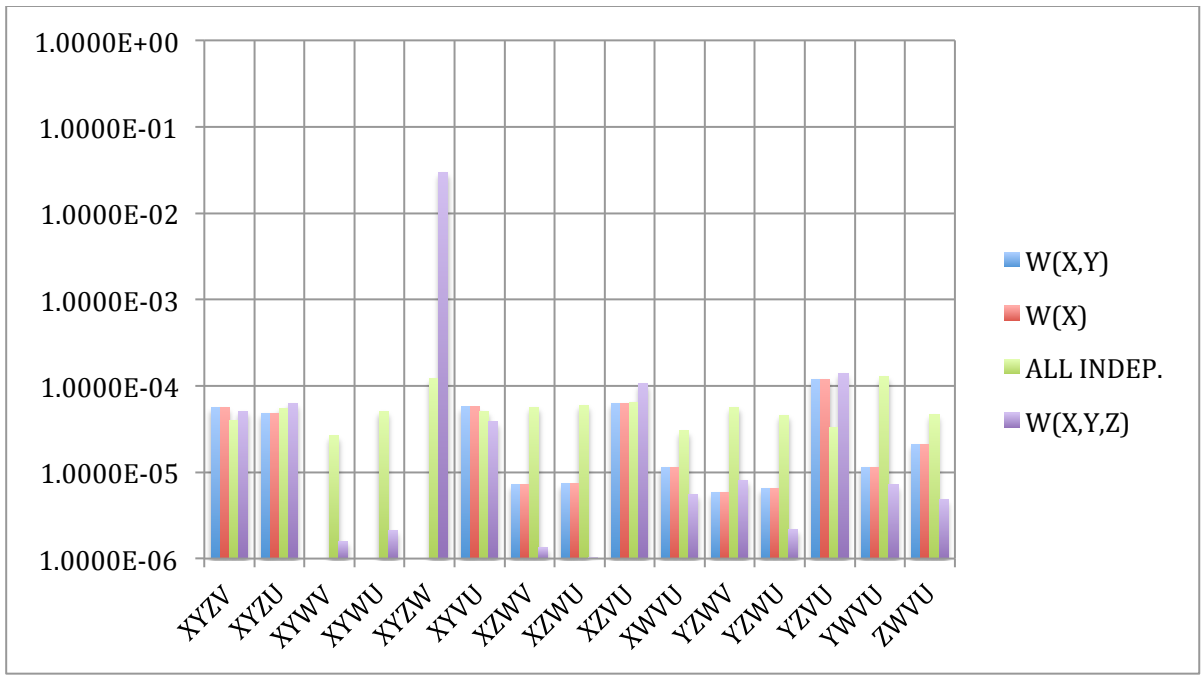

Figure 5. The four-variable measure $\overline{\Delta}$ for two data sets (1000 points, 6 variables). The dependencies are as indicated in the legend (details in Appendix)

Again, the Pearson and Spearman correlations among all pairs of variables are small (mostly <0.11, but all <0.2), but $\overline{\Lambda}$ clearly indicates the four related variables nonetheless, since it is more than two orders of magnitude greater than all others. The sensitivity to the four way dependency is indeed striking, and is not confounded by the two or three-way dependency.

**Discussion:**

A useful representation and mathematical description of the degrees of complexity of complex systems like machines, economies, biological cells and organisms is a significant challenge. It is a challenge that is at the heart of the realization of systems biology into a systematic, quantitative approach to biology. We previously approached this problem focused on biological information by defining a context-dependent measure based on pair-wise relationships [1]. In this paper we present a broad generalization for an arbitrary number of variables, which is the basis for a self-consistent, general, descriptive theory of the complexity of systems that are described by the dependencies among many variables. Our approach to multi-variable systems revolves around the question of how we can describe the joint probability density of the $n$ variables that define the system. The characterization of the joint probability density distribution is at the heart of describing the mathematical dependency among the variables, but the use of information theory measures is more forgiving of sampling limitations than direct estimates of probability densities, since many probability densities yield similar or the same information measures. A major property of our generalization is that the multivariable symmetric measure is non-zero only if the multiple variables are truly collectively dependent. It is not confounded by dependence among subsets, but it can sometimes be zero in the presence of collective dependence (sometimes termed "conditional independence".)

This theory can represent a very broad scope of systems, but our focus has been on biological complexity. The pair-wise measure presented previously [1] is the two-variable limit of our general measure. We reviewed the information approach to multi-variable dependency that relies on the concept of interaction information [5-9], and defined a new measure, which we call <u>differential interaction information</u>, that has a number of useful properties. At the three variable level it is equivalent to the conditional mutual information. We were able to make the connection between the "set complexity" of a set of variables, as previously defined for two variables in [1] and differential interaction information, by focusing on the limiting constraints we previously defined. Set complexity, originally devised as a measure of biological information, has been used as a tool to analyze genetic data [11] and to examine and describe the complexity of graphs [12,13], but is limited in its capacity for more complex systems descriptions.

We propose that our general complexity measure, or series of measures, provides a deeper and more sophisticated definition of "biological information," and is also applicable to any complex system. The differential interaction information is the central quantity in this theory, and it has a number of interesting properties, particularly its limiting values at extremes, that makes it both a useful data analysis tool as well as being a natural connection with set complexity.

We propose that the description of complex systems properties with hypergraphs can represent a great deal of the important internal dependencies that generate the complexity of the system. The multi-variable differential interaction information provides a tool that can be used to calculate the weights of the hypergraph edges given a data set describing the system [4]. We outline here a general scheme for this analysis, which is applicable in principle to any data set (figure 6.)

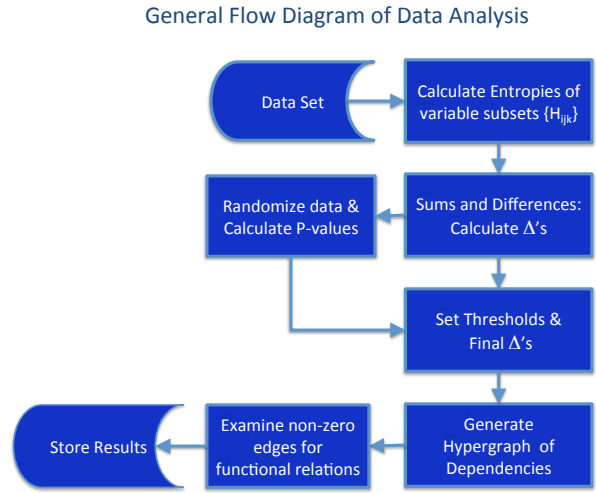

Figure 6.  A general flow diagram illustrating the analysis of data sets to the level of weighted hypergraph generation.

If we limit the number of variables to be included in the subsets, the number of marginal entropies that need to be calculated is not too large, but it is this combinatorial explosion of quantities to be calculated that will ultimately be limiting for very large data sets.  This is a well recognized practical problem in information measure calculations.  Various heuristics can be used, however, to limit the range of computed quantities, and this problem has been examined in some detail in entropy calculations.  The symmetrized form of this differential measure is remarkably sensitive to multivariable dependencies as demonstrated above.  While the calculations are difficult the extreme sensitivity suggests that approximating heuristics may retain sufficient sensitivity to allow significant computational shortcuts.  This will be explored in future work.

Note, however, that this sensitivity does not imply that the form or character of the dependency, the function itself, can be determined by these methods.  The two problems, of sensitive detection of dependence, and the characterization of the dependence function itself, are quite distinct problems and can be separated to good effect. For the complex functions like those used in our simulations (see Appendix) their characterization is a very difficult problem requiring much more data than we have provided, but the detection of dependence can be accomplish with this amount of data. This separation of problems also suggests that some of the approximating algorithms for marginal entropies may retain significant sensitivity while reducing computational complexity. We will consider the dependence function problem in future work.

The previous sections provide the means to infer the structure of a hypergraph description of a system represented by a data set and its dependencies. If we are given a data set consisting of a set of values of $n$ variables, we can then carry out the steps defined in the flowchart of figure 6. This diagram provides a general approach to the calculation of the complexities of systems represented by complex data sets, and to the inference of a hypergraph representation. Such a complex set of dependencies is indicated in figure 7 (only non-zero $\overline{\Delta_i}$'s are shown) illustrating the complexity that can be represented in this way.

Equation 14 provides a general formulation of the measure for dependence. In the product we *unbias* the choice of the additional variables since the product is invariant under permutation of the variables. The expression provides for the calculation of all of the differential interaction information measures from the marginal entropies.

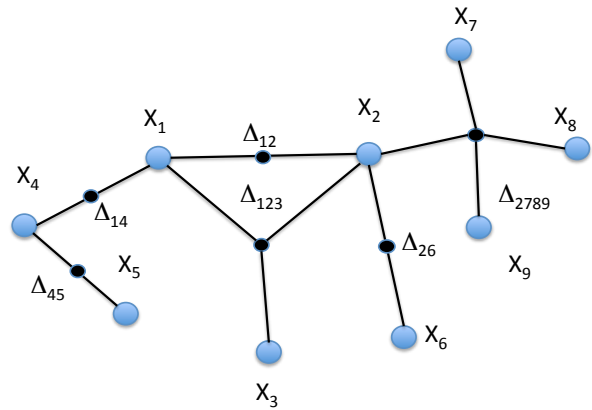

Figure 7. The complex of dependencies among a set of 9 variables (nodes) is illustrated in a hypergraph. This is same structure as in Figure 1 showing the non-zero $\overline{\Delta}(\tau_m)$'s associated with the hyperedges (in this case the subscripts indicate the nodes connected, the dependency variables.)

There are some cases where the choice of a target variable is clear and the product form of the measure (equation 14) is unnecessary. Genetic analysis is a case in point. In this case we have a large number of genetic markers as variables, with a limited number of alternative values (usually two) and usually a single target variable that is the phenotype. We classify the dependence, and the measure, in this case, as "asymmetric." In the case where we look for pair-wise synthetic genetic effects, for example, we examine all instances of pairs of markers ($X_1, X_2$) determining the phenotype ($Y$).

In addition to the calculation of the hypergraph edge weights, described above, we can consider the necessary self-consistency of the edge weights. These relations will allow the minimal hypergraph description of the dependencies implied by the data in question. This topic will be an important component of the effort to discover the causal models that are consistent with the data, but the analysis of self-consistency relations is beyond the scope of the current paper. It will be dealt with in a future publication.

In examining dependencies among variables we will, of course, want to know the nature of the dependencies, which is not addressed here. Our approach provides a theoretical context and a method to discover the existence of dependencies, consistent with the data, that can be further investigated to create sets of hypotheses. We emphasize again the distinction between detection of dependency and discovering information about the nature of the dependency. The latter can be thought of as hypothesis testing. Related to the question of consistent hypotheses, it is also natural to ask about specific causality among real variables. While the characterization of the joint probability distribution is insufficient to address the issue of causality, the changes in associations among variables as indicated by $\overline{\Delta}$ with different subsets of variables and different data sets may be able to contribute to causality indicators. We suggest that while the present theory is limited to the probability calculus, it can be useful in providing some language to extend the theory into the realm of causal tests [16,17]. The central idea here is that with the appropriate assumptions, the effect on the symmetric measure of dropping a specific variable from consideration should be able to be used in a calculus of causality of that variable. The introduction of directionality in the hypergraphs could be used to extend the present theory into the realm of representation of causality models as discussed by Pearl [17].

Finally, it is important to note, as we showed in equation 12, that the $\Delta$ specific to three variables is equivalent to conditional mutual information. Conditional mutual information has been used previously by other researchers in analyzing biological systems. Califano and colleagues have applied this measure in the analysis of gene regulatory systems [18-22]. Our work generalizes this measure for an arbitrary number of variables, and makes the connection to measures of complexity and biological information.

In summary, we have woven together three important concepts in dealing with dependency in real complex systems: measures of complexity based on information theory ideas that are applicable to biology and similarly complex systems generalizing our previous work, multi-variable dependency, and the representation of systems by hypergraphs. We expect that the methods described here can be extended further, but are already powerful in extracting patterns and dependencies from large data sets of all kinds, and in understanding the complexity of the systems reflected by data sets.

**Acknowledgements:**

We thank Hong Qian, Joseph Nadeau and Stuart Kauffman for comments on an earlier version of the manuscript, and anonymous referees for useful suggestions. We are grateful for funding provided by the Luxembourg Centre for Systems Biomedicine (LCSB) of the University of Luxembourg and the U.S. National Science

Foundation, IIS-134619, with additional support provided by the Pacific Northwest Diabetes Research Institute.

## Appendix:   Simulated data

We generated sets of data (1000 points each) with different dependencies among the variables, from complete independence to four variable dependence.

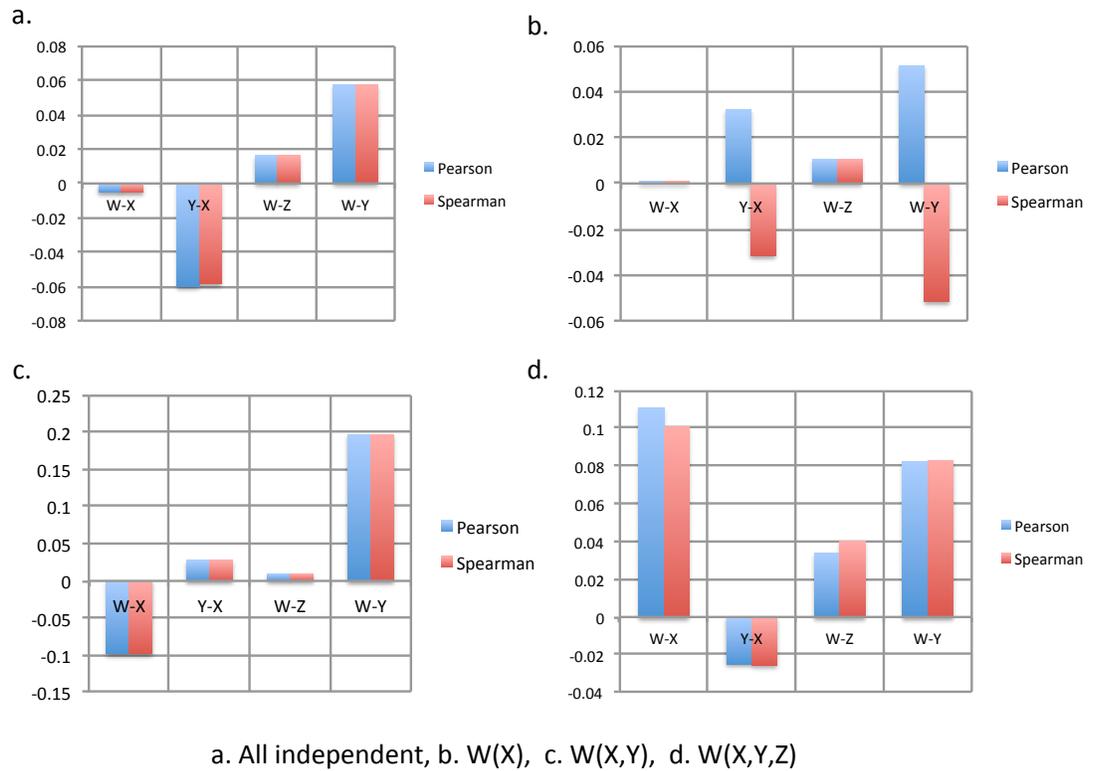

a. All independent, b. W(X),  c. W(X,Y),  d. W(X,Y,Z)

Figure A1. The Pearson and spearman correlation calculated for several pairs of variables in the 1000 point simulated set.  The functions used to generate the dependencies for b and c are shown below.

The data were generated with all variables but W random (X,Y,Z,U,V).  In case a, W is also an independent random variable, whereas in cases b-d the dependence of W on X, Y, and Z was specified according to the tables below.

| X | W(X) |
|---|---|
| 0 | 1 |
| 1 | 3 |
| 2 | 0 |

| 3 | 2 |
|---|---|

**W(X,Y)**

| X/Y | 0 | 1 | 2 | 3 |
|-----|---|---|---|---|
| 0 | 1 | 3 | 2 | 1 |
| 1 | 3 | 0 | 0 | 3 |
| 2 | 2 | 0 | 1 | 2 |
| 3 | 1 | 0 | 3 | 2 |

**W(X,Y,Z)**

| X | Y | Z | W | X | Y | Z | W |
|---|---|---|---|---|---|---|---|
| 0 | 0 | 0 | 3 | 2 | 0 | 0 | 0 |
| 0 | 0 | 1 | 0 | 2 | 0 | 1 | 0 |
| 0 | 0 | 2 | 0 | 2 | 0 | 2 | 0 |
| 0 | 0 | 3 | 3 | 2 | 0 | 3 | 1 |
| 0 | 1 | 0 | 0 | 2 | 1 | 0 | 0 |
| 0 | 1 | 1 | 3 | 2 | 1 | 1 | 0 |
| 0 | 1 | 2 | 0 | 2 | 1 | 2 | 1 |
| 0 | 1 | 3 | 1 | 2 | 1 | 3 | 1 |
| 0 | 2 | 0 | 0 | 2 | 2 | 0 | 0 |
| 0 | 2 | 1 | 0 | 2 | 2 | 1 | 1 |
| 0 | 2 | 2 | 0 | 2 | 2 | 2 | 1 |
| 0 | 2 | 3 | 1 | 2 | 2 | 3 | 1 |
| 0 | 3 | 0 | 3 | 2 | 3 | 0 | 1 |
| 0 | 3 | 1 | 1 | 2 | 3 | 1 | 1 |
| 0 | 3 | 2 | 1 | 2 | 3 | 2 | 1 |
| 0 | 3 | 3 | 3 | 2 | 3 | 3 | 1 |
| 1 | 0 | 0 | 0 | 3 | 0 | 0 | 3 |
| 1 | 0 | 1 | 3 | 3 | 0 | 1 | 1 |
| 1 | 0 | 2 | 0 | 3 | 0 | 2 | 1 |
| 1 | 0 | 3 | 1 | 3 | 0 | 3 | 3 |
| 1 | 1 | 0 | 3 | 3 | 1 | 0 | 1 |
| 1 | 1 | 1 | 2 | 3 | 1 | 1 | 2 |
| 1 | 1 | 2 | 0 | 3 | 1 | 2 | 1 |
| 1 | 1 | 3 | 2 | 3 | 1 | 3 | 2 |
| 1 | 2 | 0 | 0 | 3 | 2 | 0 | 1 |
| 1 | 2 | 1 | 0 | 3 | 2 | 1 | 1 |
| 1 | 2 | 2 | 1 | 3 | 2 | 2 | 1 |
| 1 | 2 | 3 | 1 | 3 | 2 | 3 | 1 |
| 1 | 3 | 0 | 1 | 3 | 3 | 0 | 3 |
| 1 | 3 | 1 | 2 | 3 | 3 | 1 | 2 |
| 1 | 3 | 2 | 1 | 3 | 3 | 2 | 1 |
| 1 | 3 | 3 | 2 | 3 | 3 | 3 | 2 |

Table A1.  Dependency Tables of the variable W, on others as shown: W(X), W(X,Y) and W(X,Y,Z).

To illustrate the function W(X,Y,Z) we treat the three variables as the digits of a mod4 number, in order to show the function as a two dimensional plot. The function is shown in figure A2.

a.

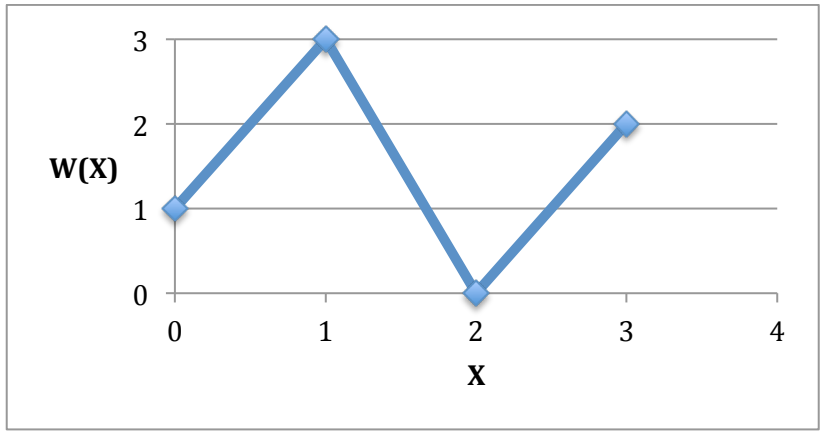

b.

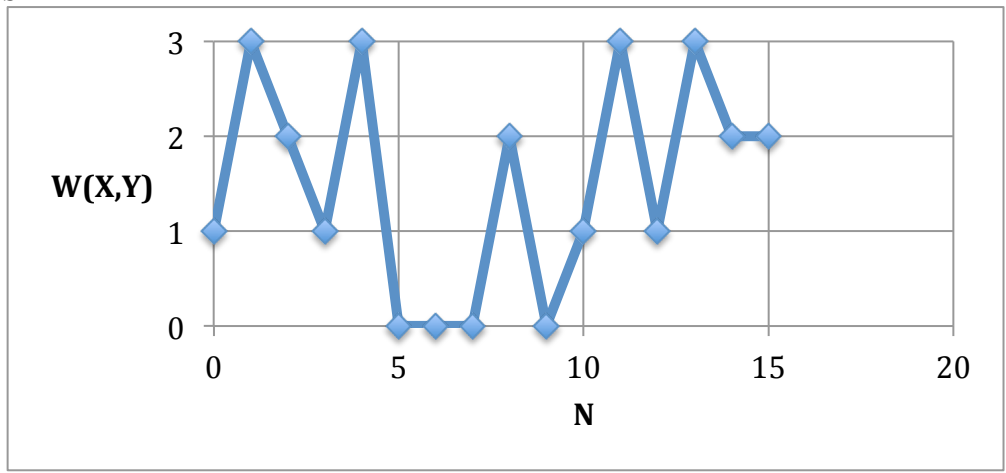

c.

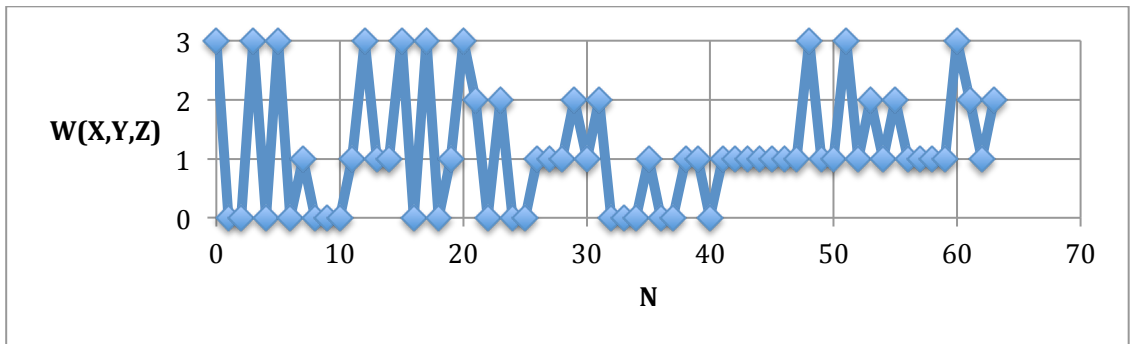

Figure A2. The functions defining the dependent variable W: W(X), W(X,Y) W(X,Y,Z). The variables X, Y and Z are the digits of a mod4 number N (depicted as a mod10 number

here.)  a. W(X), b. W(X,Y)  is  shown as a function of  N = X+4Y.  c.  W(X,Y,Z) is  shown as a function of  N = X+4Y+16Z.

While the complex and apparently arbitrary functions are such that several combinations of variables yield the same value of W (for W(X,Y) and W(X,Y,Z)) , the important point is that the other variables determine the value of W, and that the dependence of W on the other variables is complete.

**The Effects of Sample Size and Noise on the Dependence Measure**

We now consider the set of six variables, {X, Y, Z, W, U, V }, where variable W is a function of X and Y (three variable dependency), while all the rest of the variables are independent. We generated a large set D of 5000 samples and used this data set to examine the effects of sample size and noise on the symmetric delta measure. The following examples are studied in the context of the general problem of finding a triplet of interdependent variables in the presence of other independent variables. In this case we have 20 variable triplets, and we expect only one, XYW, to have a large non-zero delta value.

We first examine the fluctuations as a function of data sampling.  We consider a partition of D into M equal subsets, $D_i \subset D$, $\cup_i D_i = D$, $D_i \cap D_j = \varnothing$.  We considered two cases:  50 subsets with 100 samples each, and 10 subsets with 500 samples each.   For each subset, 20 symmetric delta values were computed corresponding to the 20 triplets.   We then computed two means/standard deviations:  one of delta for XYW across all M subsets and the other one of delta across all M subsets and all remaining 19 triplets. Table A2 shows the results.  In both cases (when the subset size equals 100 or 500) the symmetric delta of XYW is considerably lower than the average value of symmetric delta of other triplets.  As expected, the results are better when the sample size is larger (500).   While random resampling has a small effect on the values of symmetric delta at this sample size range, we are able to clearly distinguish XYW from the other triplets.

| Size | $mean(\overline{\Delta}^a)$ | $std(\overline{\Delta}^a)$ | $mean(\overline{\Delta}^b)$ | $std(\overline{\Delta}^b)$ |
|------|------|------|------|------|
| 100 | -1.8429 | 0.2757 | -0.0367 | 0.0196 |
| 500 | -2.0554 | 0.1107 | -0.0003 | 0.0002 |

Table A2: The behavior of $\overline{\Delta}$ on samples of equal size data sets. Mean and standard deviation of symmetric delta for XYW ($\overline{\Delta}^a$) averaged across all subsets, as well as mean and standard deviation of symmetric delta averaged across all other 19 triplets and all subsets ($\overline{\Delta}^b$).

How does the number of samples affect the value of symmetric delta? We consider an initial subset, $D_0 \subset D$, of 100 samples, and then incrementally construct 49 more subsets by adding 100 more samples each time, such that $D_{49} = D$. We then computed the symmetric delta for all 20 triplets in each subset. Figure A3 shows that $\overline{\Delta}$ of XYW is considerably different from $\overline{\Delta}$ of other triplets. On the other hand, this difference increases as the size of an underlying subset increases to 300 and more. It is clear that in the absence of noise the dependence in our example is detected very well with 200 to 300 samples.

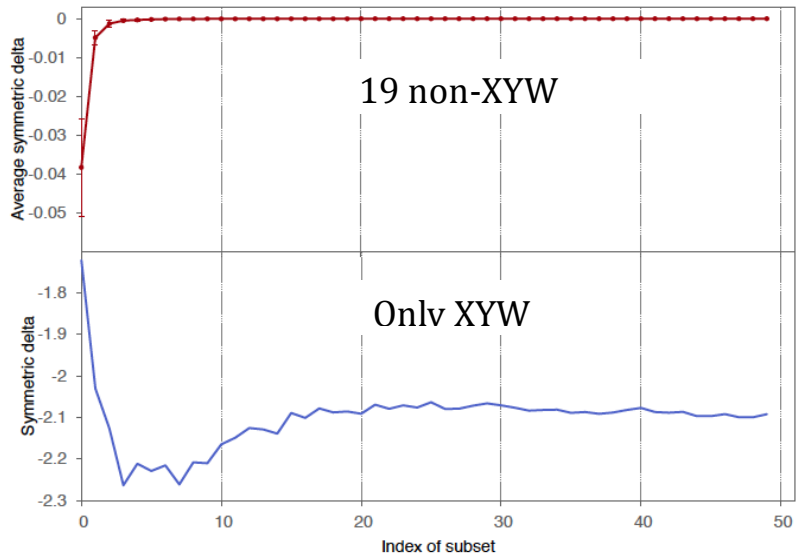

Figure A3: Behavior of $\overline{\Delta}$ on increasing size of data sets. (top) Average $\overline{\Delta}$ across 19 triplets other than XYW. The error bars correspond to standard deviation. (bottom) $\overline{\Delta}$ of XYW. The x-axis corresponds to the index $i$ of the subset Di, such that its size is $|Di| = 100 * (i + 1)$.

Finally, we investigate how the amount of noise affects the value of symmetric delta. In this example we only use a data set of 500 samples. We define a noise level parameter of a random variable as a number of samples chosen at random that are flipped to different value for that variable. We consider 20 noise levels, starting from 25 flipped values (5% noise) and ending with 500 flipped values (100% noise) each time increasing the number of flipped samples by 25. For each noise level we construct 10 data sets with random positions and values of the flips.

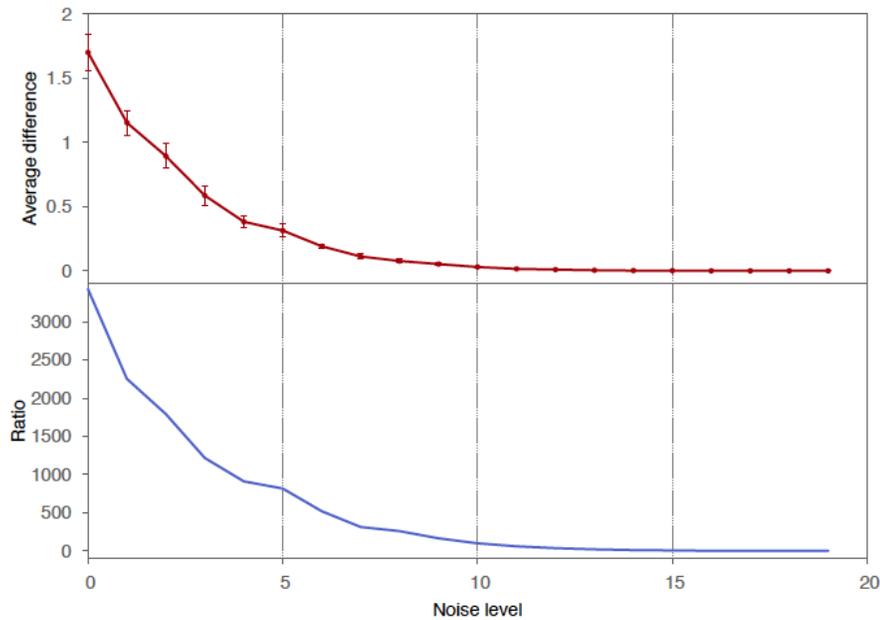

Figure A4: Behavior of $\overline{\Delta}$ on data sets with increasing noise level. (top) Average difference between $\overline{\Delta}$ of XYW and all other 19 triplets across 19 triplets. The error bars correspond to standard deviation. (bottom) Ratio of average $\overline{\Delta}$ of XYW and average $\overline{\Delta}$ of other 19 triplets. The x-axis corresponds to the index $i$ of the subset noise level, such that the number of flipped samples in a set is 25(i+1). The maximum at 20 corresponds to 100% of the 500 being subjected to random flips.

Both the average difference and the ratio between $\overline{\Delta}$ of XYW and other triplets is illustrated in figure A4. The difference was computed between $\overline{\Delta}$ of XYW and $\overline{\Delta}$ of every other triplet for every set of 10 (1900 combinations for each noise level), and the mean and standard deviation computed. The ratio was computed between $\overline{\Delta}$ of XYW averaged across 10 sets and $\overline{\Delta}$ of all other triplets averaged across all triplets and 10 sets.  The dependence on the noise level is approximately exponential.